\documentclass{llncs}

\usepackage{amsmath}
\usepackage{graphicx}
\usepackage{xcolor}

\usepackage{hyperref}
\usepackage{xspace}
\usepackage{arydshln}

\usepackage{amssymb}

\usepackage{listings}
\makeatletter
\def~{\ifmmode\;\else\penalty\@M\ \fi}
\def\@setmcodes#1#2#3{{\count0=#1 \count1=#3
 \loop \global\mathcode\count0=\count1 \ifnum \count0<#2
 \advance\count0 by1 \advance\count1 by1 \repeat}}
\DeclareSymbolFont{italic}{OT1}{\rmdefault}{m}{it}
\let\mathit\undefined
\DeclareSymbolFontAlphabet{\mathit}{italic} \edef\@tempa{\hexnumber@\symitalic}
\@setmcodes{`A}{`Z}{"7\@tempa41} \@setmcodes{`a}{`z}{"7\@tempa61} \makeatother

\newcommand{\keyword}[1]{\mbox{\lstinline[language=OOSC2Eiffel]|#1|}}
\newcommand{\classname}[1]{\keyword{#1}\normalsize}
\newcommand{\boogie}[1]{\mbox{\lstinline[language=Boogie]|#1|}}

\newcommand{\AutoProof}{AutoProof\xspace}

\newif\ifdraft\drafttrue

\ifdraft
	\newcommand\comm[1]{
		\marginpar{\raggedright\hbadness=10000\scriptsize\it #1\par}}
\else
	\newcommand\comm[1]	{}
\fi

\begin{document}

\mainmatter              

\title{Verifying Eiffel Programs with Boogie}
\titlerunning{Verifying Eiffel Programs}  
%

\toctitle{Verifying Eiffel Programs with Boogie}

\author{Julian Tschannen \and Carlo A. Furia \and Martin Nordio \and Bertrand Meyer}
\tocauthor{Tschannen, Furia, Nordio, Meyer}
\authorrunning{Tschannen et al.}   

\institute{Chair of Software Engineering, ETH Zurich, Switzerland\\
\email{\{firstname.lastname\}@inf.ethz.ch}
}

\maketitle

\begin{abstract}
Static program verifiers such as Spec\#, Dafny, jStar, and VeriFast define the state of the art in automated functional verification techniques.
The next open challenges are to make verification tools usable even by programmers not fluent in formal techniques.
This paper presents \AutoProof, a verification tool that translates Eiffel programs to Boogie and uses the Boogie verifier to prove them. 
In an effort to be usable with real programs, \AutoProof fully supports several advanced object-oriented features including polymorphism, inheritance, and function objects.
\AutoProof also adopts simple strategies to reduce the amount of annotations needed when verifying programs (e.g., frame conditions).
The paper illustrates the main features of \AutoProof's translation, including some whose implementation is underway, and demonstrates them with examples and a case study. 
\end{abstract}


\section{Usable Verification Tools}\label{intro}


It is hard to overstate the importance of \emph{tools} for software verification: tools have practically demonstrated the impact of general theoretical principles, and they have brought automation into significant parts of the verification process.
Program provers, in particular, have matured to the point where they can handle complex properties of real programs.
For example, provers based on Hoare semantics---e.g., Spec\#~\cite{BarnettLeinoSchulte05} and ESC/Java~\cite{FlanaganLeinoETAL02}---support models of the heap to prove properties of object-oriented applications; other tools using separation logic---e.g., jStar~\cite{DistefanoParkinson08} and VeriFast~\cite{JacobsSmansPiessens}---can reason about complex usages of pointers, such as in the visitor, observer, and factory design patterns.
The experience gathered so far has also outlined some design principles, which buttress the development on new, improved verification tools; the success of the Spec\# project, for example, has shown the value of using intermediate languages (Boogie~\cite{Leino08}, in the case of Spec\#) to layer a complex verification process into simpler components, which can then be independently improved and reused across different projects.

The progress of verification tools is manifest, but it is still largely driven by challenge problems and examples.
While case studies will remain important, verification tools must now also become more practical and \emph{usable} by ``lay'' programmers.
In terms of concrete goals, prover tools should support the complete semantics of their target programming language; they should require minimal annotational effort besides writing ordinary pre and postconditions of routines (methods); and they should give valuable feedback when verification fails.

The present paper describes \AutoProof, a static verifier for Eiffel programs that makes some progress towards these goals of increased usability.
\AutoProof translates Eiffel programs annotated with contracts (pre and postconditions, class invariants, intermediate assertions) into Boogie programs.
The translation currently handles sophisticated language features such as exception handling and function objects (called \emph{agents} in Eiffel parlance, and \emph{delegates} in C\#).
To reduce the need for additional annotations, \AutoProof includes simple syntactic rules to generate standard frame conditions from postconditions, so that programmers have to write down explicit frame conditions only in the more complex cases.

This paper outlines the translation of Eiffel programs into Boogie, focusing on the most original features such as exception handling (which is peculiarly different in Eiffel, as opposed to other object-oriented languages such as Java and C\#) and the generation of simple frame conditions.
The translation of more standard constructs is described elsewhere~\cite{Tsc09-Julian}.
At the time of writing, \AutoProof does not implement the translation of exceptions described in the paper, but its inclusion is underway.
The paper also reports a case study where we automatically verify several Eiffel programs, exercising different language features, with \AutoProof.
\AutoProof is part of EVE (Eiffel Verification Environment), the research branch of the EiffelStudio integrated development environment, which integrates several verification techniques and tools.
EVE is distributed as free software and freely available for download at:
\url{http://se.inf.ethz.ch/research/eve/}

\textbf{Outline.} Section~\ref{exceptions} presents the Boogie translation of Eiffel's exception handling primitives; Section~\ref{inheritance} describes a translation of conforming inheritance that supports polymorphism; Section~\ref{translation} sketches other features of the translation, such as the definition of ``default'' frame conditions; Section~\ref{applications} illustrates the examples verified in the case study; Section~\ref{relatedWork} presents the essential related work, and Section~\ref{conclusions} outlines future work.

\section{Exceptions} \label{exceptions}
Eiffel's exception handling mechanism is different than most object-oriented programming languages such as C\# and Java. 
This section presents Eiffel's mechanism (Section~\ref{sec:eiff-except-mech}), discusses how to annotate exceptions (Section~\ref{sec:spec-except}), and describes the translation of Eiffel's exceptions to Boogie (Section~\ref{sec:translation-boogie}) with the help of an example (Section~\ref{sec:example}).

\subsection{How Eiffel Exceptions Work} \label{sec:eiff-except-mech}
Eiffel exception handlers are specific to each routine, where they occupy an optional \keyword{rescue} clause, which follows the routine body (\keyword{do}).
A routine's \keyword{rescue} clause is ignored whenever the routine body executes normally. 
If, instead, executing the routine body triggers an exception, control is transferred to the  \keyword{rescue} clause for exception handling.
The exception handler will try to restore the object state to a condition where the routine can execute normally.
To this end, the body can run more than once, according to the value of an implicit variable \keyword{Retry}, local to each routine: when the execution of the handler terminates, if \keyword{Retry} has value \keyword{True} the routine body is run again, otherwise \keyword{Retry} is \keyword{False} and the pending exception propagates to the \keyword{rescue} clause of the caller routine.

Figure~\ref{fig-exceptions-eiffel} illustrates the
Eiffel exception mechanism with an example. 
The routine \keyword{attempt_transmission} tries to transmit a message by calling \keyword{unsafe_transmit}; if the latter routine terminates normally, 
 \keyword{attempt_transmission} also terminates normally without executing the \keyword{rescue} clause. On the contrary, an exception triggered by \keyword{unsafe_transmit}
 transfers control to the \keyword{rescue} clause, which re-executes the body for \keyword{max_attempts} times; if all the attempts fail to execute successfully, the attribute (field) \keyword{failed} is set and the exception propagates. 
\begin{figure}[th]
\small
\lstset{language=OOSC2Eiffel}
\begin{lstlisting}
attempt_transmission (m: STRING)
    local
       failures: INTEGER
    do
    		failed := False
        unsafe_transmit (m)
    rescue
        failures := failures + 1
        if failures < max_attempts then
            Retry := True
        else
            failed := True
        end
    end
\end{lstlisting}
\caption{An Eiffel routine with exception handler.} \label{fig-exceptions-eiffel}
\end{figure}
\normalsize

%
%

\subsection{Specifying Exceptions} \label{sec:spec-except}
The postcondition of a routine with \keyword{rescue} clause specifies the program state both after normal termination and when an exception is triggered. 
The two post-states are in general different, hence we introduce a global Boolean variable \keyword{ExcV}, which is \keyword{True} iff the routine has triggered an exception.
Using this auxiliary variable, specifying postconditions of routines with exception handlers is straightforward.
For example, the postcondition of routine \lstinline|attempt_transmission| in Figure~\ref{fig-exceptions-eiffel} says that \keyword{failed} is \keyword{False} iff the routine executes normally:
\small
\lstset{language=OOSC2Eiffel}
\begin{lstlisting}[escapechar=\#]
attempt_transmission (m: STRING)
    ensure
        ExcV implies failed
        not ExcV implies not failed
\end{lstlisting} 
\normalsize
%


The example also shows that the execution of a \keyword{rescue} clause behaves as a loop: a routine \lstinline|r| with exception handler \lstinline|r do $s_1$ rescue $s_2$ end| behaves as the loop that first executes $s_1$ unconditionally, and then repeats $s_2 \,;\, s_1$ until $s_1$ triggers no exceptions or \keyword{Retry} is \keyword{False} after the execution of $s_2$ (in the latter case, $s_1$ is not executed anymore).
To reason about such implicit loops, we introduce a \emph{rescue invariant}~\cite{NordioCalcagnoMuellerMeyer09b}; the rescue invariant holds after the first execution of $s_1$ and after each execution of $s_2\,;\,s_1$.
A reasonable rescue invariant of routine \keyword{attempt_transmission} is:
\small
\lstset{language=OOSC2Eiffel}
\begin{lstlisting}[escapechar=\#]
    rescue invariant
        not ExcV implies not failed
        (failures < max_attempts) implies not failed
\end{lstlisting} 
\normalsize

\subsection{Eiffel Exceptions in Boogie} \label{sec:translation-boogie}
The auxiliary variable \keyword{ExcV} becomes a global variable in Boogie, so that every assertion can reference it.
The translation also introduces an additional precondition \boogie{ExcV = false} for every translated routine, because normal calls cannot occur when exceptions are pending, and adds ExcV to the modifies clause of every procedure.
Then, a routine with body $s_1$ and rescue clause $s_2$ becomes in Boogie:
\small
\lstset{language=Boogie}
\begin{lstlisting}[escapechar=\#,numbers=none]
    #$\nabla(s_1, excLabel)$#
    excLabel: while (ExcV) 
              invariant #$\nabla(I_{rescue})$#;
                {
                  ExcV := false;
                  Retry := false;
                  #$\nabla(s_2, endLabel)$#  
                  if (!Retry) {ExcV := true; goto endLabel} ;
                  #$\nabla(s_1, excLabel)$#
                }
    endLabel:     
\end{lstlisting} 
\normalsize
where $\nabla(s,l)$ denotes the Boogie translation $\nabla(s)$ of the instruction $s$, followed by a jump to label $l$ if $s$ triggers an exception:
\begin{equation*}
\nabla(s, l) \ =\ \begin{cases}
  \nabla(s', l)\,;\,\nabla(s'', l) &  \text{if } s \text{ is the compound }s'\,;\,s'' \\
  \nabla(s) \,;\, \text{\lstinline[language=Boogie]|if (ExcV) $\,$\{goto l;\}|}  &  \text{otherwise}
 \end{cases}
\end{equation*}

Therefore, when the body $s_1$ triggers an exception, \lstinline|ExcV| is set and the execution enters the rescue loop. 
On the other hand, an exception that occurs in the body of $s_2$ jumps out of the loop and to the end of the routine.

The exception handling semantics is only superficially similar to having control-flow breaking instructions such as \emph{break} and \emph{continue}---available in languages other than Eiffel---inside standard loops: the program locations where the control flow diverts in case of exception are implicit, hence the translation has to supply a conditional jump after every instruction that might trigger an exception.
This complicates the semantics of the source code, and correspondingly the verification of Boogie code translating routines with exception handling. 

\subsection{An Example of Exception Handling in Boogie} \label{sec:example}

Figure~\ref{fig-exceptions-boogie} shows the translation of the example in Figure~\ref{fig-exceptions-eiffel}. 
To simplify the presentation, Figure~\ref{fig-exceptions-boogie} renders the attributes \keyword{max_attempts}, \keyword{failed}, and \keyword{transmitted} (set by \keyword{unsafe_transmit}) as variables rather than locations in a heap map.
The loop in lines 22--36 maps the loop induced by the \keyword{rescue} clause, and its invariant (lines 23--24) is the rescue invariant.


\begin{figure}
\small
\lstset{language=Boogie}
\begin{lstlisting}[escapechar=\#,numbers=left]
  var max_attempts: int; var failed:bool; var transmitted:bool;

  procedure unsafe_transmit (m: ref);
        free requires ExcV ==  false;
        modifies ExcV, transmitted;
        ensures  ExcV <==> ! transmitted;

  procedure attempt_transmission (m: ref);
        free requires ExcV ==  false;
        modifies ExcV, transmitted, max_attempts, failed;
        ensures  ExcV <==> failed;

  implementation attempt_transmission (m: ref)
  {
    var failures: int;
    var Retry: bool;  
    entry:
        failures := 0; Retry := false;
        failed := false;
        call unsafe_transmit (m); if (ExcV) { goto excL; }       
        excL:
            while (ExcV) 
                invariant ! ExcV ==> ! failed;
                invariant (failures < max_attempts) ==> ! failed;
            {    
                ExcV := false; Retry := false;
                failures := failures + 1; 
                if (failures < max_attempts) {
                    Retry := true;
                } else {
                    failed := true;
                }
                if (! Retry) {ExcV := true; goto endL;}    
                failed := false
                call unsafe_transmit (m); if (ExcV) { goto excL; }
            }
        endL: return;
  }     
\end{lstlisting} 
\caption{Boogie translation of the Eiffel routine in Figure~\ref{fig-exceptions-eiffel}.} \label{fig-exceptions-boogie}
\end{figure}
\normalsize 

\section{Inheritance and Polymorphism} \label{inheritance}
The redefinition of a routine $r$ in a descendant class can \emph{strengthen} $r$'s original postcondition by adding an \keyword{ensure then} clause, which conjoins the postcondition in the precursor.
The example in Figure~\ref{example-exp} illustrates a difficulty occurring when reasoning about postcondition strengthening in the presence of polymorphic types.
The deferred (abstract) class \classname{EXP} models nonnegative integer expressions and provides a routine \lstinline|eval| to evaluate the value of an expression object; even if \lstinline|eval| does not have an implementation in \classname{EXP}, its postcondition specifies that the evaluation always yields a nonnegative value stored in attribute \mbox{\lstinline|last_value|,} which is set as side effect (see Section \ref{sec:frame_inference}).
Classes \classname{CONST} and \classname{PLUS} respectively specialize \classname{EXP} to represent integer (nonnegative) constants and addition.
Class \classname{ROOT} is a client of the other classes, and its \lstinline|main| routine attaches an object of subclass \classname{CONST} to a reference with static type \classname{EXP}, thus exploiting polymorphism.

\lstset{language=OOSC2Eiffel}
\begin{figure}[th]
\begin{center}
\begin{tabular}{p{.40\textwidth} p{.55\textwidth}}
\begin{lstlisting}[escapechar=\#]
deferred class EXP
feature
    last_value: INTEGER
    eval
        deferred
        ensure
            last_value >= 0
        end
end
 #\ # 
 #\ # 
 #\ # 
\end{lstlisting}
&

\begin{lstlisting} 
class PLUS inherit EXP feature 
    left, right: EXP
    eval do
        		left.eval; right.eval
            last_value := left.last_value + 
                          right.last_value
        ensure then
            last_value = left.last_value + 
                         right.last_value
        end	
invariant
    no_aliasing: left /= right /= Current
end
\end{lstlisting} \\


\begin{lstlisting}
class CONST inherit EXP
feature 
    value: INTEGER
    eval
        do
            last_value := value
        ensure then
            last_value = value
        end
invariant
    positive_value: value >= 0
end
\end{lstlisting}
&

\begin{lstlisting}[escapechar=\#]
class ROOT
feature    
    main
        local
        		e: EXP
        do
            e := create {CONST}.make (5);
            e.eval
            check e.last_value = 5 end    
        end
end 
\end{lstlisting}
\\

\end{tabular}
\end{center}
\caption{Nonnegative integer expressions.}\label{example-exp}
\end{figure}

The verification goal consists in proving that, after the invocation $e.eval$ (in 
class \classname{ROOT}), \mbox{\lstinline|eval|'s} postcondition in class \classname{CONST} holds, which subsumes the \lstinline|check| statement in the caller.
Reasoning about the invocation only based on the static type \lstinline|EXP| of the target \lstinline|e| guarantees the postcondition \lstinline|last_value >= 0|, which is however too weak to establish that \lstinline|last_value| is exactly 5.

Other approaches, such as M\"uller's~\cite{Mueller02}, have targeted these issues in the context of Hoare logics, but they usually are unsupported by automatic program verifiers.
In particular, with the Boogie translation of polymorphic assignment implemented in Spec\#, we can verify the assertion \lstinline|check e.last_value = 5 end| in class \classname{ROOT} only if \lstinline|eval| is declared \emph{pure}; \lstinline|eval| is, however, not pure.
The Spec\# methodology selects the pre and postconditions according to static types for non-pure routines: the call \lstinline|e.eval| only establishes \lstinline|e.last_value >= 0|, not the stronger \linebreak\lstinline|e.last_value = 5| that follows from \lstinline|e|'s dynamic type \lstinline|CONST|, unless an explicit cast redefines the type \lstinline|CONST|.
The rest of the section describes the solution implemented in \AutoProof, which handles contracts of redefined routines.



\subsection{Polymorphism in Boogie}
The Boogie translation implemented in \AutoProof can handle polymorphism appropriately even for non-pure routines; it is based
on a methodology for agents~\cite{NordioCalcagnoMeyerMuellerTschannen10} and on a methodology for pure routines~\cite{DarvasLeino07,LeinoMueller08}.
The rest of the section discusses how to translate postconditions of redefined routines in a way that accommodates polymorphism, while still supporting modular reasoning.
Eiffel also supports \emph{weakening of preconditions} in redefined routines; the translation to Boogie handles it similarly as for postconditions (we do not discuss it for brevity).


The translation of the postcondition of a routine $r$ of class $X$ with result type $T$ (if any) relies on an auxiliary function \boogie{post.X.r}:
\small
\begin{lstlisting}[language=Boogie,escapechar=\#,numbers=none]
  function post.X.r (h1, h2: HeapType; c: ref; res: T) returns (bool);
\end{lstlisting}
\normalsize
which predicates over two heaps (the pre and post-states in $r$'s postcondition), a reference \boogie{c} to the current object, and the result \boogie{res}.
$r$'s postcondition in Boogie references the function \boogie{post.X.r}, and it includes the translation $\nabla_{post}(X.r)$ of $r$'s postcondition clause syntactically declared in class \classname{X}:
\small
\begin{lstlisting}[language=Boogie,escapechar=\#,numbers=none]
  procedure X.r (Current: ref) returns (Result: T);
    free ensures post.X.r (Heap, old(Heap), Current, Result);
    ensures #$\nabla_{post}(X.r)$#
\end{lstlisting}
\normalsize
\boogie{post.X.r} is a \boogie{free ensures}, hence it is ignored when proving $r$'s implementation and is only necessary to reason about usages of $r$.

The function \boogie{post.X.r} holds only for the type $X$; for each class $Y$ which is a descendant of $X$ (and for $X$ itself), an axiom links $r$'s postcondition in $X$ to $r$'s strengthened postcondition in $Y$:
\small
\begin{lstlisting}[language=Boogie,escapechar=\#,numbers=none]
  axiom (forall h1, h2: HeapType; c: ref; r: T ::
      $\$$type(c) <: Y ==> (post.X.r(h1, h2, c, r) ==> #$\nabla_{post}(Y.r)$#));
\end{lstlisting}
\normalsize
The function \boogie{$\$$type} returns the type of a given reference, hence the postcondition predicate $post.X.r$ implies an actual postcondition $\nabla_{post}(Y.r)$ according to $c$'s dynamic type. 

In addition, for each redefinition of $r$ in a descendant class $Z$, 
the translation defines a fresh Boogie procedure $Z.r$ with 
corresponding postcondition predicate $post.Z.r$ and axioms 
for all of $Z$'s descendants.

\begin{figure}
\small
\begin{center}
\lstset{language=Boogie}
\begin{lstlisting}[escapechar=\#,numbers=left]
  function post.EXP.eval(h1, h2: HeapType; c: ref) returns (bool);

  procedure EXP.eval(current: ref);
	  free ensures post.EXP.eval(Heap, old(Heap), current, result);
	  ensures Heap[current, last_value] >= 0;
	  // precondition and frame condition omitted

  axiom (forall h1, h2: HeapType; o: ref ::
    #\$#type(o) <: EXP ==> 
      (post.EXP.eval(h1, h2, o) ==> (h1[o, last_value] >= 0)) );
  axiom (forall h1, h2: HeapType; o: ref ::
    #\$#type(o) <: CONST ==> 
      (post.EXP.eval(h1, h2, o) ==> h1[o, last_value] == h1[o, value]) );
#
%  axiom (forall h1, h2: HeapType; o: ref; r: int ::
%    h[o, $$type] <: PLUS ==> 
%      (post.EXP.eval(h1, h2, o, r) ==> (r >= 0 && 
%       r == fun.EXP.eval(h1, h1[o, left]) + fun.EXP.eval(h1, h1[o.right])))
%  );
#
  implementation ROOT.main (Current: ref) {
		  var e: ref;
    entry:
        // translation of create {CONST} e.make (5)
      havoc e;
      assume Heap[e, #\$#allocated] == false;
      Heap[e, #\$#allocated] := true;
      assume #\$#type(e) == CONST;
      call CONST.make(e, 5);
        // translation of e.eval
      call EXP.eval(e);
        // translation of check e.last_value = 5 end
      assert Heap[e, last_value] == 5;
      return;
  }
\end{lstlisting} 
\end{center}
\caption{Boogie translation of the Eiffel classes in Figure~\ref{example-exp}.} \label{fig-exp-boogie}
\end{figure}
\normalsize

\subsection{An Example of Polymorphism with Postconditions}
Figure~\ref{fig-exp-boogie} shows the essential parts of the Boogie translation of the example in Figure~\ref{example-exp}. 
The translation of routine \lstinline|eval| in lines 3--6 references the function \boogie{post.EXP.eval}; the axioms in lines 8--13 link such function to $r$'s postcondition in \classname{EXP} (lines 8--10) and to the additional postcondition introduced in \classname{CONST} for the same routine (lines 11--13).
The rest of the figure shows the translation of the client class \classname{ROOT}.

\section{Other Features}\label{translation}
This section briefly presents other features of the Eiffel-to-Boogie translation.

\subsection{Default Frame Conditions} \label{sec:frame_inference}
Frame conditions are necessary to reason modularly about heap-manipulating programs, but they are also an additional annotational burden for programmers.
In several simple cases, however, the frame conditions are straightforward and can be inferred syntactically from the postconditions.
For a routine $r$, let $mod_r$ denote the set of attributes mentioned in $r$'s postcondition; $mod_r$ is a set of (\emph{reference}, \emph{attribute}) pairs.
The translation of Eiffel to Boogie implemented in \AutoProof assumes that every attribute in $mod_r$ may be modified (that is, $mod_r$ is $r$'s frame), whereas every other location in the heap is not modified.
Since every non-pure routine already includes the whole \boogie{Heap} map in its \boogie{modifies} clause, the frame condition becomes the postcondition clause:
\small
\begin{lstlisting}[language=Boogie,escapechar=\#,numbers=none]
ensures (forall o: ref, f: Field :: (o, f) #$\notin mod_r$# ==> Heap[o, f] == old(Heap[o, f]));
\end{lstlisting}
\normalsize


To ensure soundness in the presence of inheritance, the translation always 
uses the postcondition of the original routine definition to infer the frame of 
the routine's redefinitions.

The frame conditions inferred by \AutoProof work well for routines whose postconditions only mention attributes of primitive type.
For routines that manipulate more complex data, such as arrays or lists, the default frame conditions are too coarse-grained, hence programmers have to supplement them with more detailed annotations.
Extending the support for automatically generated frame conditions is part of future work.

\subsection{Routines Used in Contracts Pure by Default} \label{pure_routines}
The translation of routines marked as \emph{pure} generates the frame condition \linebreak\mbox{\lstinline[language=Boogie]|ensures Heap == old(Heap)|} which specifies that the heap is unchanged.
\AutoProof implicitly assumes that every routine used in contracts is pure, and the translation reflects this assumption and checks its validity.
While the Eiffel language does not require routines used in contract to be pure, it is a natural assumption which holds in practice most of the times, because the behavior of a program should not rely on whether contracts are evaluated or not.
Therefore, including this assumption simplifies the annotational burden and makes using \AutoProof easier in practice.

\subsection{Agents}
The translation of Eiffel to Boogie supports \emph{agents} (Eiffel's name for \emph{function objects} or \emph{delegates}). 
The translation introduces abstract predicates to specify routines that take
function objects as arguments: some axioms link the abstract predicates to concrete specifications whenever an agent is initialized.
The details of the translation of agents is described elsewhere~\cite{NordioCalcagnoMeyerMuellerTschannen10}.

\section{Case Study}\label{applications}
This section presents the results of a case study applying \AutoProof to the verification of the 11 programs listed in Table~\ref{tab:examples}.
For each example, the table reports its name, its size in number of classes and lines of code, the length (in lines of code) of the translation to Boogie, the time taken by Boogie to verify successfully the example, and the kind of Eiffel features mostly exercised by the example.

Example 1 is a set of routines presented in Meyer's book~\cite{Meyer97} when describing Eiffel's exceptions; Example 2 is a set of classes part of the EiffelStudio compiler runtime. 
To verify them, we extended the original contracts with postconditions to express the behavior when exceptions are triggered, and with rescue invariants (Section~\ref{sec:spec-except}).\footnote{As the implementation in \AutoProof of translation of exceptions is currently underway, these two examples were translated by hand.}
The most difficult part of verifying these example was inventing rescue invariants. Even when the examples are simple, the rescue invariants may be subtle, because they have to include clauses both for normal and for exceptional termination. 

Examples 3--5 target polymorphism in verification. 
The \emph{Expression} example is described in Section~\ref{inheritance}. 
The \emph{Sequence} example models integer sequences with the deferred classes \classname{SEQUENCE}, 
\classname{MONOTONE_SEQUENCE}, and \lstinline|STRICT_| \lstinline|SEQUENCE|, and their effective 
descendants \classname{ARITHMETIC_SEQUENCE}, and \classname{FIBONACCI_SEQUENCE}.
The \emph{Command} example implements the command design pattern with a deferred class \classname{COMMAND} and effective descendants that augment the postcondition of \classname{COMMAND}'s deferred routine \emph{execute}.
The encoding of inheritance described in Section~\ref{inheritance} is accurate but it also significantly increases the size of the Boogie translation and correspondingly the time needed to handle it.
Since a translation that takes dynamic types into account is not always necessary, future work will introduce an option to have \AutoProof translating contracts solely based on the static type of references.


Examples 6--8 use agents and are the same examples as in~\cite{NordioCalcagnoMeyerMuellerTschannen10}. The \emph{Formatter} example
illustrates the specification of functions taking agents as arguments; the
\emph{Archiver} example uses an agent with closed arguments; the \emph{Calculator} example implements the command design pattern using agents rather than subclasses. 

Examples 9--11 combine multiple features: a cell class
that stores integer values; a counter that can be
increased and decreased; a bank account class with clients.
These examples demonstrate other features of the translation, such as 
the usage of default frame conditions.

The source code of the examples is available at
\url{http://se.ethz.ch/people/tschannen/boogie2011_examples.zip}. 
The experiments ran on a Windows~7 machine with a 2.71 GHz dual core 
Intel Pentium processor and 2GB of RAM.

\begin{table}[t]
\begin{tabular}{ l l  c c c c c }
		 & \textsc{Example name} \ \ & \textsc{Classes}\ \ & \textsc{LOC Eiffel}\ \ \ \ & \textsc{LOC Boogie}\ \ \ \ & \textsc{Time [s]} & \textsc{Feature} \\ \hline
  1. &Textbook OOSC2          & 1 & 106 & 481 & 2.33 & Exceptions  \\
  2. &Runtime ISE                  & 4 & 203 & 561 & 2.32 & Exceptions  \\
  	\hdashline[1pt/5pt]
  3. & Expression & 4 & 134 & 752 & 2.11 & Inheritance \\ 
  4. & Sequence & 5 & 195 & 976 & 2.28 & Inheritance \\
  5. & Command & 4 & 99 & 714 & 2.14 & Inheritance \\
  	\hdashline[1pt/5pt]
  6. & Formatter & 3 & 120 & 761 & 2.23 & Agents \\
  7. & Archiver & 4 & 121 & 915 & 2.07 & Agents \\
  8. & Calculator & 3 & 245 & 1426 & 9.73 & Agents \\
  	\hdashline[1pt/5pt]
  9. & Cell / Recell & 3 & 154 & 905& 2.09 & General \\
  10. & Counter & 2 & 97 & 683 & 2.02 & General \\
  11. & Account & 2 & 120 & 669 & 2.04 & General \\
  	\hline
     & \textbf{Total} & \textbf{35} & \textbf{1594} & \textbf{8843} & 31.36 &  \\
     \ \\
\end{tabular}

\caption{Examples automatically verified with \AutoProof}
\label{tab:examples}
\end{table}

\section{Related Work}\label{relatedWork}
Tools such as ESC/Java~\cite{FlanaganLeinoETAL02} and
Spec\#~\cite{BarnettLeinoSchulte05} have made considerable progress towards practical and automated functional verification. 
Spec\# is an extension of C\# with syntax to express preconditions, postconditions, class invariants, and non-null types. 
Spec\# is also a verification environment that verifies Spec\# programs by translating them to Boogie---also developed within the same project. 
Spec\# works on significant examples, but it does not support every feature of C\# (for example, delegates are not handled, and exceptions can only be checked at runtime). 
Spec\# includes annotations to specify frame conditions, which make proofs easier but at the price of an additional annotational burden for
developers. To ease the annotational overhead, Spec\# adds a default frame condition that includes all attributes of the target object. This solution has the advantage that the frame can change with routine redefinitions to include attributes introduced in the subclasses.
\AutoProof follows a different approach and tries to rely on standard annotations whenever possible, which impacts on the programs that can be verified automatically.

Spec\# has shown the advantages of using an intermediate language for
verification. 
Other tools such as Dafny~\cite{Leino10} and Chalice~\cite{LeinoMueller09},
and techniques based on Region Logic~\cite{BanerjeeNaumannRosenberg08}, follow this approach, and they also rely on Boogie as intermediate language and verification back-end, in the same way as \AutoProof does.

Separation logic~\cite{OHearnYangReynolds04} is an extension of Hoare logic with connectives that define separation between regions of the heap, which provides an elegant approach to reasoning about programs with mutable data structures.
Verification environments based on separation logic---such as jStar~\cite{DistefanoParkinson08} and VeriFast~\cite{JacobsSmansPiessens}---can verify advanced features such as usages of the visitor, observer, and
factory design patterns.  
On the other hand, writing separation logic annotations requires
considerably more expertise than using standard contracts embedded in the
programming language; this makes tools based on separation logic more
challenging to use by practitioners.

\section{Future Work}\label{conclusions}
\AutoProof is a component of EVE, the Eiffel Verification Environment, which combines different verification tools to exploit their synergies and provide a uniform and enhanced usage experience, with the ultimate goal of getting closer to the idea of ``verification as a matter of course''.

Future work will extend \AutoProof and improve its integration with other verification tools in EVE.
In particular, the design of a translation supporting the expressive model-based contracts~\cite{PFM10-VSTTE10} is currently underway. 
Other aspects for improvements are a better inference mechanism for frame conditions and intermediate assertions (e.g., loop invariants~\cite{FM10-YG70-post}); a support for interactive prover as an alternative to Boogie for the harder proofs; and a combination of \AutoProof with the separation logic prover also part of EVE~\cite{StadenCM10}.

\bibliographystyle{abbrv}
\bibliography{literatur}


\end{document}